\documentclass[
    ,final            % use final for the camera ready runs
  ]
  {aipproc}

\layoutstyle{6x9}

\begin{document}

\title{Effects of Helium Phase Separation on the Evolution of Giant Planets}

\author{Jonathan J. Fortney}{
  address={Lunar and Planetary Laboratory, Department of Planetary Sciences,\\The University of Arizona, \\1629 E. University Blvd., Tucson, AZ 85721}
}

\author{William B. Hubbard}{
  address={Lunar and Planetary Laboratory, Department of Planetary Sciences,\\The University of Arizona, \\1629 E. University Blvd., Tucson, AZ 85721}
}

%%%%%%%%%%%%%%%%%%%%%%%%%%%%%%%%%%%%%%%%%%%%%%%%%%%%%%%%%%%%

\begin{abstract}
We present the first models of Saturn and Jupiter to couple their evolution to both a radiative-atmosphere grid and to high-pressure phase diagrams of hydrogen with helium. The purpose of these models is to quantify the evolutionary effects of helium phase separation in Saturn's deep interior.  We find that prior calculated phase diagrams in which Saturn's interior reaches a region of predicted helium immiscibility do not allow enough energy release to prolong Saturn's cooling to its known age and effective temperature. We explore modifications to published phase diagrams that would lead to greater energy release, and find a modified H-He phase diagram that is physically reasonable, leads to the correct extension of Saturn's cooling, and predicts an atmospheric helium mass fraction $Y_{atmos}$ in agreement with recent estimates.  We then expand our inhomogeneous evolutionary models to show that hypothetical extrasolar giant planets in the 0.15 to 3.0 Jupiter mass range may have $T_{eff}$s 10-15 K greater than one would predict with models that do not incorporate helium phase separation.
\end{abstract}

\maketitle

%%%%%%%%%%%%%%%%%%%%%%%%%%%%%%%%%%%%%%%%%%%%
%% MAINMATTER
%%%%%%%%%%%%%%%%%%%%%%%%%%%%%%%%%%%%%%%%%%%%

\section{Introduction}

The interiors of Jupiter and Saturn, extrasolar giant planets (EGPs), and brown dwarfs (BDs) are all described by similar physics: these bodies are mainly composed of liquid metallic hydrogen, and their interior energy transport mainly occurs through efficient convection \cite{Hubbard02}, leading to largely isentropic interiors.  Jupiter and Saturn, whose radius, mass, luminosity, and age are known precisely, can serve as calibrators of thermal-history calculations for the entire class of objects.  They can provide a test of the adequacy of the diverse physical models, including interior thermodynamics, heat transport mechanisms, and model-atmosphere grid, that enter into the general thermal-history theory for EGPs and BDs.  However, at very low effective temperatures ($\sim$~100 K), the corresponding interior temperatures may become low enough for phase separation of abundant interior components to occur, and this effect must be quantitatively evaluated before Jupiter and Saturn can be used as calibrators.  This work provides a quantitative assessment of inhomogeneous evolution in Jupiter, Saturn, and low-mass EGPs.

\noindent Relevant calculations and data for Jupiter and Saturn follow:
\begin{itemize}
\item Saturn is currently over 50\% more luminous than one would predict with a homogeneous, adiabatic cooling model.  Saturn models reach the planet's known $T_{eff}$ of 95.0 K in only 2.0 to 2.5 Gyr \cite{FH03}.
\item For Jupiter, homogeneous, adiabatic cooling models allow Jupiter to reach its known $T_{eff}$ of 124.4 K in $\sim$~4.7 Gyr.  This is a good match.
\item The atmospheres of both Jupiter and Saturn are depleted in helium relative to the expected protosolar helium mass fraction of $\sim 0.27$.  $Y_{atmos}$ for Jupiter is 0.231 \cite{vonzahn98} and for Saturn is likely 0.18-0.25 \cite{CG00}, but is poorly constrained.
\end{itemize}

\section{Hydrogen and Helium under High Pressure}

\begin{figure}
  \includegraphics[height=.3\textheight]{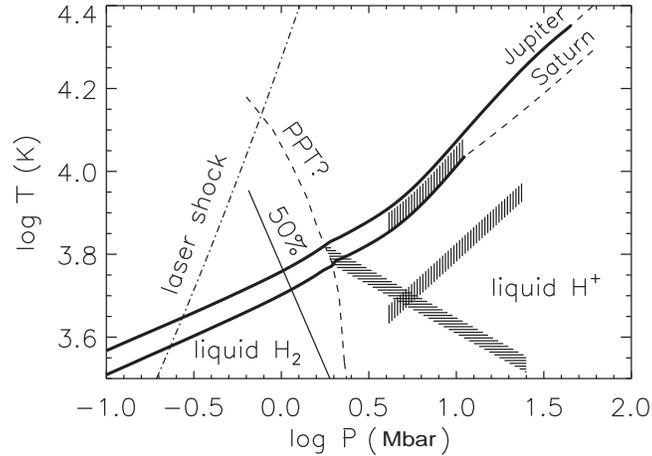}
  \caption{Temperature-pressure plot of the interiors of Jupiter and Saturn at $t$ = 4.56 Gyr, superimposed on a hydrogen phase diagram (see text for details).  The upper boundary of the horizontally-hatched region shows the minimum temperature at which He is fully miscible in metallic hydrogen with a mass fraction $Y=0.27$, while the lower boundary shows the minimum temperature corresponding to $Y=0.21$, according to HDW \cite{HDW} theory.  The lower vertically-hatched region shows the same He miscibility limits according to Pfaffenzeller, et al.~\cite{Pfaff}, while the upper vertically-hatched region shows the modification to the Pfaffenzeller et al. theory that gives a realistic prolongation of Saturn's age.  With this phase diagram, when He rains down, it falls all the way down to the core.  Liquid metallic hydrogen (H$^+$), liquid molecular hydrogen (H$_2$), and the maximum pressures and temperatures reached by laser shock data are also shown.
\label{figure:p1}}
\end{figure}

Under extreme pressure dense molecular hydrogen dissociates and ionizes to become liquid metallic hydrogen.  The transition is likely continuous over a pressure of 1-5 Mbar at the temperatures of interest ($\sim 10^4$ K) in giant planets.  Also at these temperatures, helium, which makes up about 8\% of the atoms in a solar composition mixture, likely becomes immiscible in liquid metallic hydrogen \cite{Stevenson75,HDW,Pfaff}.  Figure~\ref{figure:p1} shows the interior temperature-pressure profile of Jupiter and Saturn superimposed on a hydrogen phase diagram.

Detailed calculations on the dynamics and distribution of helium in giant planets have been performed \cite{SS77b}.  These authors found that when helium becomes immiscible in liquid metallic hydrogen, the composition that separates out is essentially pure helium, and this helium on fairly short timescales (relative to the convective timescale) will coalesce to form helium droplets.  These droplets, being denser than the surrounding liquid metallic hydrogen, will fall through the planet's gravitational field.  If the droplets reach a region where helium is again miscible at higher concentration, they will redissolve, enriching the deeper regions of the planet in helium.  Helium would be lost from $all$ regions with pressures lower than the pressures in the immiscibility region, since the planet is fully convective up to the atmosphere.  Excess helium would be mixed down to the immiscibility region and be lost to deeper layers.  This ``helium rain'' could be a substantial additional energy source for giant planets.  

\section{Calculations}
\begin{figure}
  \includegraphics[height=.34\textheight]{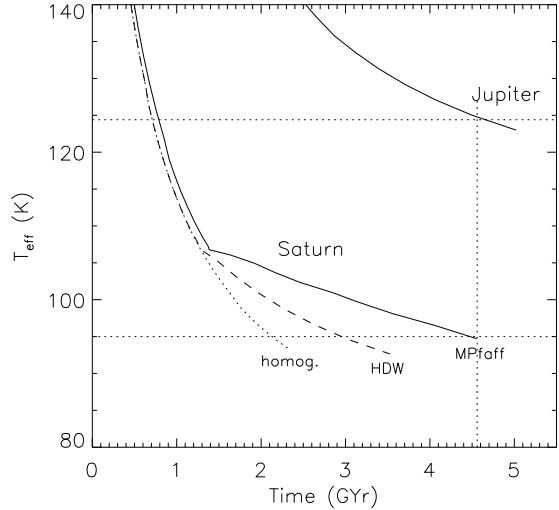}
  \caption{Evolutionary models including the phase separation of helium from liquid metallic hydrogen.  The dotted curve for Saturn and the solid curve for Jupiter are for homogeneous evolution.  The dashed curve includes the phase diagram of HDW\cite{HDW}, while the solid curve for Saturn uses the proposed modified phase diagram (MPfaff).  The modified phase diagram allows both Jupiter and Saturn to reach their known ages and $T_{eff}$s.  The interior of Jupiter begins He phase separation at $t \approx 5$ Gyr.
\label{figure:js2}}
\end{figure}

\begin{figure}
  \includegraphics[height=.38\textheight]{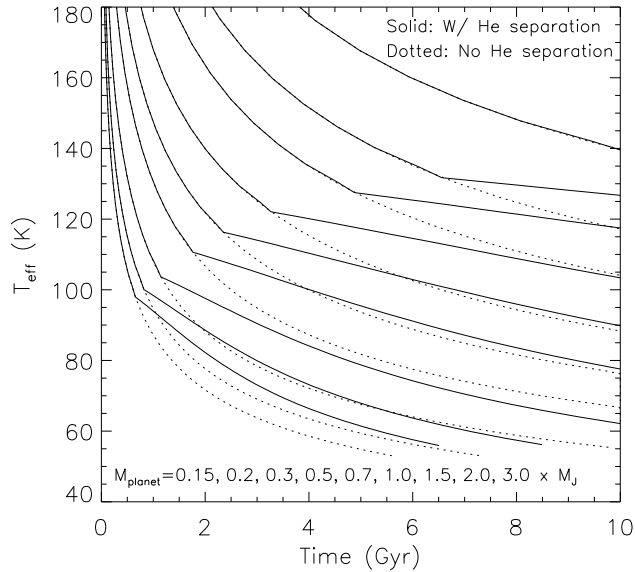}
  \caption{Evolution of the $T_{eff}$ for planets of mass 0.15 to 3.0 $M_{\rm J}$ for our standard models with no stellar irradiation and 10 $M_{\rm Earth}$ cores.  The dotted lines are models without helium phase separation, while the solid lines include the effects of helium phase separation on the planets' cooling.  The top curve is for the highest mass planet while the bottom curve is for the lowest mass planet.
\label{figure:Teff}}
\end{figure}

We find that the calculated phase diagram of HDW \cite{HDW}, which is essentially equivalent to that of Stevenson \cite{Stevenson75}, are inapplicable to the interiors of giant planets, if helium phase separation is Saturn's only additional energy source.  As \mbox{Figure~\ref{figure:js2}} shows, this phase diagram prolongs Saturn's cooling 0.8 Gyr, even in the most favorable circumstance that all energy liberated is available to be radiated, and does not instead go into heating the planet's deep interior.  As we show in our published work \cite{FH03}, we find that a modified version of the phase diagram of Pfaffenzeller et al. \cite{Pfaff}, with a higher temperature for the onset of helium immiscibility (see Figure~\ref{figure:p1}), allows Saturn to reach its known $T_{eff}$ and age, while Jupiter evolves homogeneously until $t \approx$ 5 Gyr.  Saturn's $Y_{atmos}$ falls to 0.185 at 4.5 Gyr, which is at the low end of Saturn's derived value \cite{CG00}.
This modified phase diagram can be applied to various hypothetical giant planets \cite{FH04}.  We follow the evolution of planets with masses from 0.15 $M_{\rm J}$ (1/2 Saturn's mass) to 3.0 $M_{\rm J}$.  Figure~\ref{figure:Teff} shows the evolution of these planets with and without the effects of helium phase separation.  These planets are in isolation and possess 10 $M_{\rm Earth}$ cores.  At Gyr ages the model planets undergoing phase separation can have $T_{eff}$s 10-15 K higher than the homogeneous models, making the planets somewhat easier to detect.
\begin{theacknowledgments}
We thank Adam Burrow, Dave Sudarksy, \& Tristan Guillot for interesing conversations.  JJF is funded by a NASA GSRP Fellowship and WBH by a NASA PG\&G grant.
\end{theacknowledgments}

\bibliographystyle{aipproc}

\end{document}